\documentclass{article}
\usepackage{fullpage}
\usepackage{amsmath,amssymb}
\usepackage{amsthm}
\usepackage{algorithmic,algorithm2e}
\usepackage{graphicx}
\usepackage{tikz}
\usetikzlibrary{calc}
\usepackage[left=1in,top=1in,right=1in]{geometry}
\usepackage{subfigure}
\usepackage{color}
\usepackage[noblocks]{authblk}

\newtheorem{theorem}{Theorem}
\newtheorem{lemma}{Lemma}
\newtheorem{proposition}{Proposition}
\newtheorem{corollary}{Corollary}

\newcommand{\ZZ}{{\mathbb Z}}

\tikzstyle{black node}=[draw,circle,fill=black, minimum size=3pt, inner sep=0pt]
\tikzstyle{blue node}=[draw,circle,fill=cyan!30!white, minimum size=3pt, inner sep=0pt]
\tikzstyle{red node}=[draw,circle,fill=red!75!black, minimum size=3pt, inner sep=0pt]
                            
\tikzstyle{edge} = [draw,thin,-,black]
\tikzstyle{nonedge} =  [draw,-,black,dashed]
\tikzstyle{thick nonedge} =  [draw,thick,-,black,dashed]
\tikzstyle{thick edge} = [draw,thick,black]

\tikzstyle{obedge} = [draw,thick,-,fill=green]
\tikzstyle{obedge2} = [draw,line width=2pt,-,green]
\tikzstyle{smallob} = [draw,rectangle,fill=green, minimum size = 10pt]
\tikzstyle{vsmallob} = [draw,rectangle,fill=green, minimum size = 5pt]

\tikzstyle{carve} = [draw,yellow!20,fill=yellow!20]
\tikzstyle{carve2} = [draw,black,fill=gray]
\tikzstyle{carveob} = [draw,black,fill=gray]
\tikzstyle{carvewhite} = [draw,white,fill=white]

\title{Convex Obstacle Numbers of Outerplanar Graphs\\and Bipartite Permutation Graphs\thanks{Radoslav Fulek and Noushin Saeedi's research was supported by Swiss National Science Foundation grant 200021-125287/1 and the Bernoulli Center at EPFL.  Noushin Saeedi's research was performed during a visit to EPFL during which she was partially supported by EPFL.  Deniz Sar{\i}{\"o}z's research was supported by NSA grant 47180-0001.}}

\author{Radoslav Fulek}
\affil{\'Ecole Polytechnique F\'ed\'erale de Lausanne\authorcr
{\tt radoslav.fulek@epfl.ch}
\authorcr
\ }

\author{Noushin Saeedi}
\affil{The University of British Columbia\authorcr
{\tt noushins@cs.ubc.ca}
\authorcr
\ }

\author{Den{\.{i}}z Sar{\i}{\"{o}}z}
\affil{The Graduate School and University Center of The City University of New York\authorcr
{\tt sarioz@acm.org}
\authorcr
\ }

\begin{document}

\maketitle
\thispagestyle{empty} 
\begin{abstract}
The disjoint convex obstacle number of a graph $G$ is the smallest number $h$ 
such that there is a set of $h$ pairwise disjoint convex polygons (obstacles) and a set of $n$ points in the plane (corresponding to $V(G)$) so that a vertex pair $uv$ is an edge if and only if the corresponding segment $\overline{uv}$ does not meet any obstacle.

We show that the disjoint convex obstacle number of an outerplanar graph is always at most 5,
and of a bipartite permutation graph at most 4. 
The former answers a question raised by Alpert, Koch, and Laison.
We complement the upper bound for outerplanar graphs with the lower bound of 4.
\end{abstract}
\section{Introduction and Preliminaries}
\label{sec:intro}
An \emph{obstacle representation} of a graph $G$, as first defined by Alpert, Koch, and Laison \cite{alpert}, is a straight-line drawing of $G$, together with a set of polygonal obstacles such that two vertices of $G$ are connected with an edge if and only if the line segment between the corresponding points does not meet any of the obstacles. As they did, 
we assume the points corresponding to the graph vertices together with the polygon vertices are in general position (no three on a line).
An obstacle representation of $G$ with $h$ obstacles is called an $h$-\emph{obstacle representation} of $G$. The \emph{obstacle number} of $G$ is the smallest number of obstacles needed in an obstacle representation of $G$. If we restrict the polygonal obstacles to be convex, we call such a representation a \emph{convex obstacle representation}. \emph{Convex obstacle number} and $h$-\emph{convex obstacle representation} are defined similarly.

If the convex obstacles are required to be pairwise disjoint, we call such a representation a \emph{disjoint convex obstacle representation}, and define 
\emph{disjoint convex obstacle number} and
$h$-\emph{disjoint convex obstacle representation} similarly.
Surely, the convex obstacle number of a graph is at most its disjoint convex obstacle number.
We conjecture that there are graphs having convex obstacle number strictly less than their disjoint convex obstacle number, 
so we reason about these two parameters separately.

In \cite{obstacleWG}, it was shown that for any fixed $h$, the number of graphs on $n$ (labeled) vertices with obstacle number at most $h$ is at most 
$2^{O(hn\log^2 n)}$.
From this, it follows that every graph class with $2^{\omega(n \log ^2 n)}$ members on $n$ vertices
(such as the class of all bipartite graphs) has unbounded obstacle number.
It was also shown therein that the number of unlabeled graphs on $n$ vertices with convex obstacle number at most $h$ is at most
$2^{O(hn \log n)}$.
Since the number of planar graphs on $n$ vertices is $2^{\Theta(n \log n)}$ 
(see \cite{numberPlanar} for exact asymptotics), 
the bounds given by \cite{obstacleWG} are inconclusive regarding the obstacle number or convex obstacle number of 
the class of planar graphs or a subclass.

Nonetheless, it was shown by Alpert, Koch, and Laison \cite{alpert} that every outerplanar graph admits a 1-obstacle representation in which the obstacle is in the unbounded face. 
The same paper raised the question of whether the convex obstacle number of an outerplanar graph can be arbitrarily large.
We answer this question in negative.
In particular, we prove the following two results regarding outerplanar graphs in Sections \ref{sec:outerplanar_upper_bound_5} and \ref{sec:outerplanar_lower_bound_4} respectively.

\begin{theorem}
\label{thm:outerplanar_upper_bound_5}
The convex (and disjoint convex)
obstacle number 
of every outerplanar graph is at most \emph{five}.
\end{theorem}

\begin{theorem}
\label{thm:outerplanar_lower_bound_4}
There are trees having disjoint convex obstacle number at least \emph{four}.
\end{theorem}

In Section \ref{sec:bipartite_permutation_upper_bound_4}, we prove the following regarding bipartite permutation graphs.
\begin{theorem}
\label{thm:bipartite_permutation_upper_bound_4}
The convex (and disjoint convex) obstacle number 
of every bipartite permutation graph is at most \emph{four}.
\end{theorem}

\section[Upper bound on convex obstacle number of outerplanar graphs]{Upper bound on convex obstacle number\\of outerplanar graphs}
\label{sec:outerplanar_upper_bound_5}
\begin{proof}[Proof of Theorem \ref{thm:outerplanar_upper_bound_5}]
We shall show that the convex obstacle number of every outerplanar graph is at most \emph{five}, by giving a method to generate five convex obstacles that can represent any outerplanar graph.  
For a given connected outerplanar graph $G$, we first construct a digraph $\overrightarrow{G'}$ with certain properties, whose underlying graph is a subgraph of $G$. We call $\overrightarrow{G'}$ the BFS-digraph of $G$. We show an obstacle representation using \emph{five} convex obstacles for the BFS-digraph, and then modify the representation without changing the number of obstacles to represent the graph $G$.  We finally discuss how to accommodate the disconnected case, still with \emph{five} obstacles.

\subsection{Constructing the BFS-digraph and its properties}
\label{sec:BFS}
Let  $G$ be a connected outerplanar graph. Perform the breadth-first search based Algorithm \ref{alg:bfs-digraph} on $G$ 
that outputs a digraph which we call the BFS-digraph of $G$, and denote by $\overrightarrow {G'}$.
We say that a vertex of a {BFS-digraph} has depth $i$ if its distance from the BFS root is $i$. 

\begin{algorithm}
\KwIn {A connected graph $G = G(V,E)$}
\KwOut {The digraph $\overrightarrow G'$ called the {BFS-digraph} of $G$}
\caption{Algorithm to compute a BFS-digraph of a connected graph}
\label{alg:bfs-digraph}
\begin{algorithmic}
\STATE $V' := V_0:= $ singleton set with an arbitrarily chosen vertex of $G$ (the BFS $root$)
\STATE $\overrightarrow{E'} :=\emptyset$
\STATE $i := 0$
\WHILE{$V' \neq V$}
\STATE $V_{i+1} := \{v \mid u \in V_{i}, (u, v) \in E\} \setminus V'$
\STATE $V' := V' \cup V_{i+1}$
\STATE $\overrightarrow{E'} := \overrightarrow{E'} \cup \{\overrightarrow{(u,v)} \mid u \in V_i , v \in V_{i+1}, (u,v) \in E\}$ 
\STATE $i := i + 1$
\ENDWHILE
\RETURN $\overrightarrow G'(V, \overrightarrow{E'})$\\

\

\end{algorithmic}
\end{algorithm}

\begin{lemma}
\label{prop-of-digraph}
A BFS-digraph $\overrightarrow{G'}$ of a connected outerplanar graph $G$ has a straight-line drawing such that
\begin{enumerate}
\item each vertex at depth $i$ lies on the line $y = -i$;
\item two edges are disjoint except possibly at their endpoints; and
\item a vertical downward ray starting at a vertex $v$ meets the graph only at $v$.
\end{enumerate} 
\end{lemma}

\begin{figure}
\begin{center}
\begin{tikzpicture}[scale=1.3]
\path[draw = black] (0, 1) rectangle (11, -5); 

\newcommand{\depthmultiplier}{1.4}

\newcommand{\drawraytill}{-5}
{
	\foreach \nodname/\xcoord/\depth in {root/10/0} {
		\node[black node] (\nodname) at (\xcoord, -\depthmultiplier*\depth) [label=above:{$\nodname$}] {};
		\draw[dashed] (\nodname) -- (\xcoord, \drawraytill);
	}
}

{
	\foreach \nodname/\xcoord/\depth in {a/1/1, e/5/1, j/9/1} {
		\node[black node] (\nodname) at (\xcoord, -\depthmultiplier*\depth) [label=above left:{$\nodname$}] {};
		\draw[dashed] (\nodname) -- (\xcoord, \drawraytill);
	}
	\draw[-latex, thick] (root) -- (a);
	\draw[-latex, thick] (root) -- (e);
	\draw[-latex, thick] (root) -- (j);
}

{
	\foreach \nodname/\xcoord/\depth in {b/2/2, d/4/2, f/6/2, h/8/2} {
		\node[black node] (\nodname) at (\xcoord, -\depthmultiplier*\depth) [label=above:{$\nodname$}] {};
		\draw[dashed] (\nodname) -- (\xcoord, \drawraytill);
	}
	\draw[-latex, thick] (a) -- (b);
	\draw[-latex, thick] (a) -- (d);
	\draw[-latex, thick] (e) -- (d);
	\draw[-latex, thick] (e) -- (f);
	\draw[-latex, thick] (j) -- (h);
}

{
	\foreach \nodname/\xcoord/\depth in {c/3/3, g/7/3} {
		\node[black node] (\nodname) at (\xcoord, -\depthmultiplier*\depth) [label=left:{$\nodname$}] {};
		\draw[dashed] (\nodname) -- (\xcoord, \drawraytill);
	}
	\draw[-latex, thick] (b) -- (c);
	\draw[-latex, thick] (d) -- (c);
	\draw[-latex, thick] (f) -- (g);
	\draw[-latex, thick] (h) -- (g);
}

{
	\draw (e) -- (j);
	\draw (b) -- (d);
	\draw (f) -- (h);
}
\end{tikzpicture}
\caption{A {BFS-digraph} of an outerplanar graph $G$ drawn to exhibit the three properties in Lemma \ref{prop-of-digraph}.  The edges without arrows correspond to edges of $G$ that are not in the digraph.  
For a given outerplanar graph $G$, 
regardless of the choice of the BFS root,
there is a drawing of the resulting {BFS-digraph} that satisfies the three properties and induces a straight-line outerplanar drawing of $G$.}
\end{center}
\end{figure}
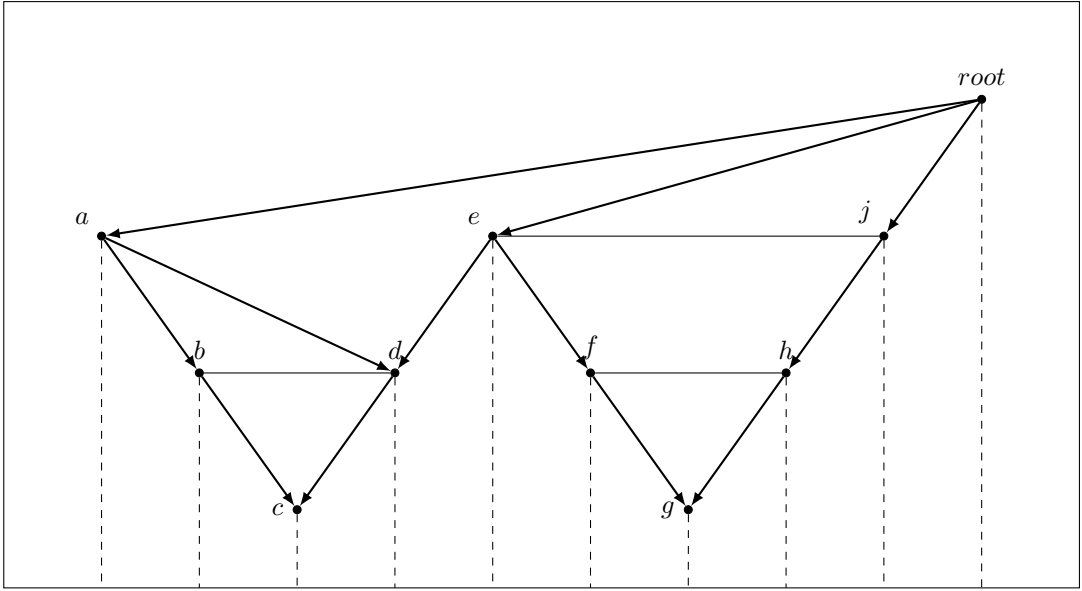

\begin{proof}
Let $\overrightarrow{G'_i}$ denote the subgraph of $\overrightarrow{G'}$ induced on vertices at depth less than or equal to $i$. We show the existence of such a drawing by constructing it. 
We will proceed by induction on {$\overrightarrow{G'_i}$}.

Consider a planar embedding of the outerplanar graph $G$ in which every vertex meets the outer face, with all vertices on a circle having {the} $root$ as its topmost point. From now on, we do not distinguish between a graph 
and its embedding.
Draw {the} $root$ on the line $y=0$. 
Then draw all its neighbors on the line $y = -1$ and to its left, preserving their order in $G$.
All arcs {corresponding to the edges between the root and its neighbors} are oriented downward. 
So far we have $\overrightarrow{G'_1}$, which satisfies the desired properties. 
We now show how to extend for $i \geq 1$
an embedding of $\overrightarrow{G'_i}$ with the desired properties to
an embedding of $\overrightarrow{G'_{i+1}}$ with the desired properties.
Let $v_{i, 1}, v_{i, 2}, \hdots, v_{i, \ell} = root$ denote the 
vertices of $\overrightarrow{G'_i}$ 
in left-to-right order.
For the sake of brevity, let $v_{i, 0}$ also denote the $root$.
The depth $i+1$ neighbors of  a vertex $v_{i,k}$ in $V_i$ 
lie in $G$ either on the clockwise arc from $v_{i,k}$ to $v_{i,k-1}$ 
or on the counterclockwise arc from $v_{i,k}$ to $v_{i,k+1}$.
Otherwise, $G$ is not planar, or its vertices are not in convex position. 
We refer to the depth $i+1$ neighbors of $v_{i,k}$ on the clockwise arc from $v_{i,k}$ to $v_{i,k-1}$ 
as the \emph{left} children of $v_{i,k}$, 
and those on the counterclockwise arc from $v_{i,k}$ to $v_{i,k+1}$ 
as the \emph{right} children of $v_{i,k}$. 
Note that
for 
vertices $v_{i,j}$ and $v_{i,j+1}$, the rightmost child of $v_{i,j}$ lies before or at the same place as the leftmost child of $v_{i,j+1}$.
We apply the following steps for each $v_{i,k}$ in $V_i$:
\begin{itemize}
\item put the left children of $v_{i,k}$, in order of clockwise proximity in $G$ to $v_{i,k}$, on the line $y= -(i+1)$ so that 
they are to the left 
of 
$v_{i,k}$ and 
(unless $k=1$) to the right of 
$v_{i,k-1}$;
\item put the right children of $v_{i,k}$, in order of counterclockwise proximity in $G$ to $v_{i,k}$, on the line $y = -(i+1)$ so that 
they are between
$v_{i,k}$ and 
$v_{i,k+1}$;
\item 
make sure that for every pair of vertices $v_{i,j}$ and $v_{i,j+1}$ in $V_i$, the rightmost child of $v_{i,j}$ and the leftmost child of $v_{i,j+1}$ preserve their order in $G$, and is embedded once if they are one and the same.
\end{itemize}
Note that due to the outerplanarity of $G$, a right descendent and a left descendant of a vertex have no common descendants, rendering the last step possible. Therefore, the extended embedding represents $\overrightarrow{G'_{i+1}}$ and satisfies all three conditions.
\end{proof}

According to this embedding, we say that two vertices are \emph{consecutive} if they are on the same horizontal line and there is no vertex between them.

\begin{corollary}
\label{cond-parents}
A vertex has at most two parents. Moreover, if a vertex $v$ has two parents, the parents are consecutive; 
and $v$ is the rightmost child of its left parent, and the leftmost child of its right parent.
\end{corollary}
\begin{proof}
If any of the conditions above does not hold, property 3 of Lemma \ref{prop-of-digraph} is violated. 
\end{proof}
By Corollary \ref{cond-parents}, we also know two vertices at depth $i$ have a common child only if they are consecutive.

\begin{corollary}
\label{right-left-child}
Two consecutive vertices such that one is a left child and the other is a right child of the same parent, do not have a common child.
\end{corollary}
\begin{proof}
It directly follows from the third property of Lemma \ref{prop-of-digraph}.
\end{proof}
 
\subsection[5-convex obstacle representation of the BFS-digraph of a connected outerplanar graph]{5-convex obstacle representation of the BFS-digraph\\of a connected outerplanar graph}
\label{sec:up:5-convex}
We demonstrate a set of five convex obstacles and describe how to place vertices of $\overrightarrow{G'}$ to obtain a 5-convex obstacle representation for $\overrightarrow{G'}$. We first describe the arrangement of the set of obstacles. We have two disjoint convex arcs symmetric about a horizontal line, such that both arcs curve toward the line of symmetry. We consider the arcs to be parts of large circles, so that they behave like lines, except that they block visibilities among vertices put sufficiently near them. In the region bounded by the two arcs, we put three line obstacles, which form an S-shape with perpendicular joints, so that the S-shape is equally far from either arc, and the projection of the S-shape onto either arc covers the whole arc. We then disconnect the line obstacles by creating a small (and similar) aperture at each joint. The arrangement of the set of obstacles is shown in Figure \ref{fig:obs-rep}.
\bigskip
\begin{figure}[htp]
\begin{center}
\includegraphics[scale=1]{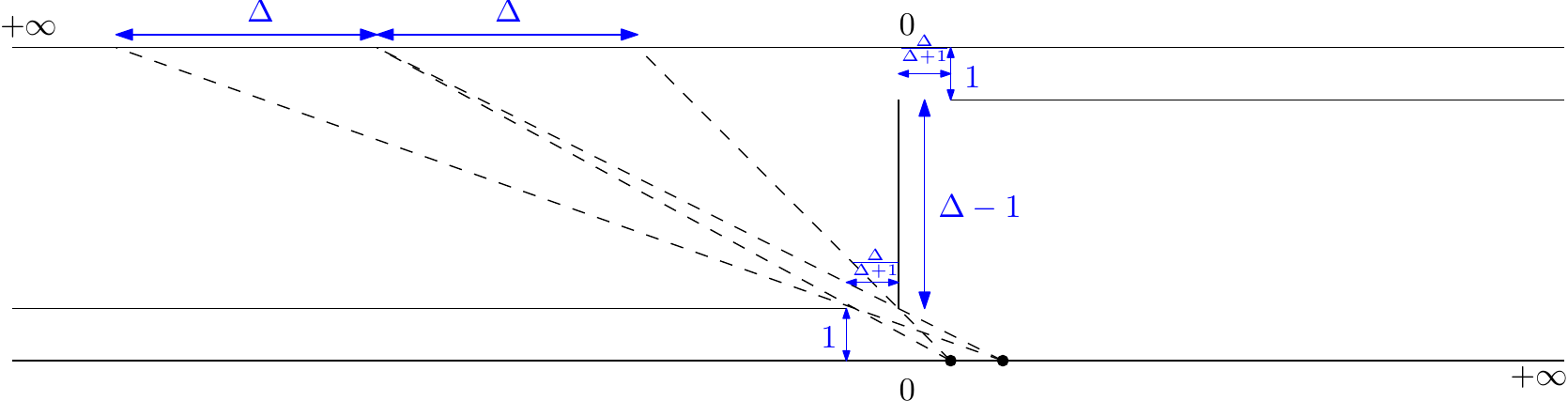}
\end{center}
\caption{The arrangement of the set of five convex obstacles.}
\label{fig:obs-rep}
\end{figure}

The key idea is to place all vertices of the graph sufficiently close to either of the arcs, and control the visibilities through the created apertures. 
For the sake of simplicity of exposition, from now on, we say a vertex is placed on an arc if it is sufficiently close to an arc. 
For each arc, the nearby and distant apertures are respectively called the \emph{{outgoing aperture}} and the \emph{{incoming aperture}}.  For each vertex on an arc, we draw the outgoing edges through the {outgoing aperture} of its underlying arc. We parameterize the arcs such that the intersection points of the extended vertical line segment of the S-shape set at the arcs mark the zeros, and the positive axes of the lower arc and the upper arc point respectively to the right and to the left.
Let {$\Delta \geq 2$} be at least the maximum outdegree in $\overrightarrow{G'}$.
We show if the S-shape is constructed so that
\begin{enumerate}
\item for two positive points unit distance apart on one arc, the parts of the opposite arc they see (through the outgoing aperture) share a single point, and
\item any point on an arc sees (through the outgoing aperture) an interval of length $\Delta$ of the other arc;
\end{enumerate}
then this obstacle set can represent BFS-digraphs of all connected outerplanar graphs.

We first investigate the structure of the set of obstacles to fulfill the conditions above.
The distance between two compact subsets of the plane is the least distance between two of their respective points. 
Denote by $w$ the aperture's width, denote by $s$ the vertical segment's length (in the S-shape), and denote by $x$ the distance between the S-shape and either arc.

Considering the arcs as lines, the first condition manifests if and only if $\frac{w}{1}=\frac{s+x}{s+2x}$, and the second condition holds if and only if $\frac{w}{\Delta}=\frac{x}{s+2x}$. 
These two equations require that $w=\frac{\Delta}{\Delta+1}$ and $s=(\Delta-1)x$, and we choose $x = 1$ to make things simple.
We next show that the depicted set of obstacles represents any BFS-digraph.
(Surely, the obstacle set depends on $\Delta$ which is conditioned on $\overrightarrow{G'}$, and to list vertex coordinates of the polygonal obstacles we would also need to know the maximum depth in $\overrightarrow{G'}$ as we will discuss, so strictly speaking we have an obstacle set \emph{template}.)

\begin{proposition}
\label{bfs-rep}
The arrangement of five convex obstacles shown in Figure \ref{fig:obs-rep}, represents BFS-digraphs of all connected outerplanar graphs.
\end{proposition}
\begin{proof}
We give an algorithm to place the vertices of a connected BFS-digraph $\overrightarrow{G'}$ so that, together with the set of obstacles, they form an obstacle representation of $\overrightarrow{G'}$. 
We consider the two arcs in the obstacle set as lines; after all vertices are placed, we curve them a bit---just as much that they block visibilities among vertices on them.  This way, we ignore visibilities among vertices on the same arc (when considered as a line) and show that the set of obstacles represents $\overrightarrow{G'}$.

Consider a drawing of $\overrightarrow{G'}$ that satisfies the conditions in Lemma \ref{prop-of-digraph}. From now on, by $\overrightarrow{G'}$ we refer to this embedding. Place the root of $\overrightarrow{G'}$ at coordinate 1 of the lower arc. We get a representation of $\overrightarrow{G'_0}$, where $\overrightarrow{G'_i}$ denotes the induced subgraph of $\overrightarrow{G'}$ containing all vertices at depth at most $i$. Suppose $\overrightarrow{G'_i}$ is represented such that
\begin{enumerate}
\item all vertices at an even depth are placed on the lower arc, and all vertices at an odd depth are placed on the upper arc;
\item on each arc, vertices at different depths are \emph{well separated}, i.e., arc intervals containing all vertices at the same depth are disjoint;
\item vertices of each depth preserve their ordering in $\overrightarrow{G'}$; and
\item every two consecutive vertices are at least one unit apart. 
\end{enumerate}  

Note that by preserving the order, we mean if a vertex is to the left of some other vertex $v$ in $\overrightarrow{G'}$, it gets a smaller coordinate than $v$ when put on an arc.

Now, we describe how to add vertices at depth $i+1$ to obtain a representation of $\overrightarrow{G'_{i+1}}$ satisfying the conditions above. Let $v_{i,j}$ denote the $j$-th vertex at depth $i$
and let $[a_{i,j},b_{i,j}]$ denote the interval of the opposite arc that is visible from $v_{i,j}$ through the {outgoing aperture}. For each vertex $v_{i,j}$ at depth $i$ in the representation of $\overrightarrow{G'_i}$, we add its children on the opposite arc as follows:
\begin{itemize}
\item If $v_{i,j}$ has a common child with its immediate preceding vertex in $V_i$, put its leftmost child at $a_{i,j}$; otherwise, put the leftmost child at $a_{i,j}+\frac{1}{2}$.
\item If $v_{i,j}$ has a common child with its immediate next vertex in $V_i$, put its rightmost child at $b_{i,j}$; otherwise, put the rightmost child at $b_{i,j}-\frac{1}{2}$.
\item Put the remaining left children, preserving their ordering, after the leftmost one so that all left children are one unit apart.
\item Put the remaining right children, preserving their ordering, before the rightmost one so that all right children are one unit apart.
\end{itemize}
Since every point sees an interval of length $\Delta$, we know $b_{i,j}=a_{i,j}+\Delta$. Thus, as each vertex has at most $\Delta$ children, by performing the above algorithm, the rightmost {left} child is placed {before} the leftmost {right child, and} are at least one unit apart. Therefore, all consecutive pairs of vertices are of distance 
at least one.
Moreover, we know every two points, which are one unit apart, have a common point-of-sight;  that is, the greatest point-of-sight of the smaller point equals the smallest point-of-sight of the greater one. By Corollaries \ref{cond-parents} and \ref{right-left-child}, we know that if two vertices have a common child, then they are consecutive; and they are not right and left children of the same parent. Therefore, the presented algorithm put vertices so that two vertices at depth $i$ and $i+1$ are visible in the representation, if and only if they are connected in $\overrightarrow{G'}$. 
Conditions 1, 3, and 4 are surely satisfied after performing the algorithm.
Since a vertex sees no other vertex except through the apertures, to complete the proof, what remains to be shown is that a vertex sees only its children through its outgoing aperture (and only its parent(s) through its incoming aperture).
To that end, next we prove that Condition 2 is satisfied, 
namely that
vertices at different depths are well separated: they lie in pairwise disjoint intervals.

Let $I_0$ denote the ``interval'' $[1,1]$ wherein the root is placed, and for every $i \geq 0$
let $I_{i+1}$ denote the interval visible from $I_{i}$ through the outgoing aperture.
Since every vertex at depth $i$ is in $I_i$, and $I_i$ and $I_{i+1}$ belong to different arcs, to prove Condition 2, it suffices to show that $I_i < I_{i+2}$ 
(i.e., every point in $I_i$ has a smaller coordinate than every point in $I_{i+2}$) for every $i \geq 0$.
If $I_i = [a,b]$, the structure of the obstacle set yields $I_{i+1} = [\Delta \times a,\Delta \times b + \Delta]$.
Since $\Delta \geq 2$, this gives $I_0 < I_2$.
By induction, we obtain that 
$I_i = [\Delta^i, 2\Delta^i + \sum_{j=1}^{i-1} \Delta^{j}]$ for every $i \geq 1$.
Since $\Delta \geq 2$, for every $i \geq 1$ we have
$2\Delta^i + \sum_{j=1}^{i-1} \Delta^{j} < 3\Delta^{i} < \Delta^{i+2}$, therefore, $I_i < I_{i+2}$.

Since we have previously ensured that a vertex $v$ at depth $i$ sees only its children through the outgoing aperture among all vertices at depth $i+1$,
the well ordering of the intervals implies that $v$ cannot see any other vertices through the outgoing aperture.
By symmetry of sight, this implies that no vertex can see through its incoming aperture any vertex other than its parent(s).

This concludes the proof that we gave an obstacle representation of 
$\overrightarrow{G'}$. 
\end{proof}

\subsection{Adjusting the representation for general outerplanar graphs}
We first show how to modify the representation of a connected BFS-digraph $\overrightarrow{G'}$ to accommodate its corresponding outerplanar graph $G$. We know that the underlying graph of $\overrightarrow{G'}$ and $G$ are the same, except that $\overrightarrow{G'}$ has no edge between two vertices at the same depth. Since $G$ is an outerplanar graph, the extra edges of $G$, if any, are such that they connect two consecutive vertices. Therefore, to allow existence of extra edges in the representation, we simply shave off the portion of the arc between their endpoints.

Now, we adapt this idea for disconnected outerplanar graphs. Let
\mbox{$C_1, C_2, \ldots, C_n$} 
be the components of a given outerplanar graph.  Let $\overrightarrow{C'_i}$ be a BFS-digraph of $C_i$, as defined in Subsection \ref{sec:BFS}.
Let $\Delta \geq 2$ be at least the maximum outdegree among all BFS-digraphs, and construct the obstacle set template as before. Now, let $L$ denote the maximum depth among all $\overrightarrow{C'_i}$. 
We declare $I_0$ to be the interval $[1,1]$ on one arc, and for every $i > 0$, we let $I_i$ be the interval $[\Delta^i, 2\Delta^i + \sum_{j=1}^{i-1} \Delta^{j}]$ on the arc opposite to interval $I_{i-1}$. The modified algorithm for representing a disconnected outerplanar graph is as follows.
For each $\overrightarrow{C'_i}$, put its root at an arbitrary place in $I_{(i-1)(L+2)}$.
Then carry out the algorithm described in Subsection \ref{sec:up:5-convex}
to place all vertices of $\overrightarrow{C'_i}$ for every $i$.
This ensures that no vertex in $C'_i$ can see a vertex of $C'_j$ for any $i \neq j$.
We then shave off the arcs as necessary to provide visibility among vertices at the same depth where desired.

We obtain a representation for an arbitrary outerplanar graph, concluding the proof of Theorem \ref{thm:outerplanar_upper_bound_5}.
\end{proof}

\section[Lower bound on disjoint convex obstacle number of outerplanar graphs]{Lower bound on disjoint convex obstacle number\\of outerplanar graphs}
\label{sec:outerplanar_lower_bound_4}

For a rooted tree, we use the standard terminology---the depth of a vertex is its topological distance to the root, and the height of the tree is the maximum depth over all its vertices.
\begin{proof}[Proof of Theorem \ref{thm:outerplanar_lower_bound_4}]
Denote by $T_{k,h}$ the full complete $k$-ary tree with height $h$ rooted at $r$.
We will show that the disjoint convex obstacle number of $T_{k,3}$ is at least \emph{four}, for $k$ to be specified later.
We say that two edges form a \emph{crossing} if they meet at an internal point of both.
(Recall that in an obstacle representation, no three vertices are collinear.)

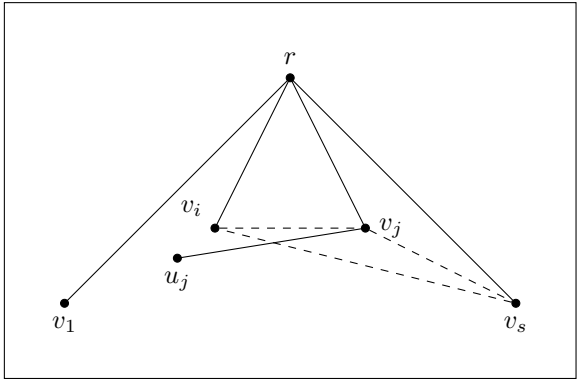
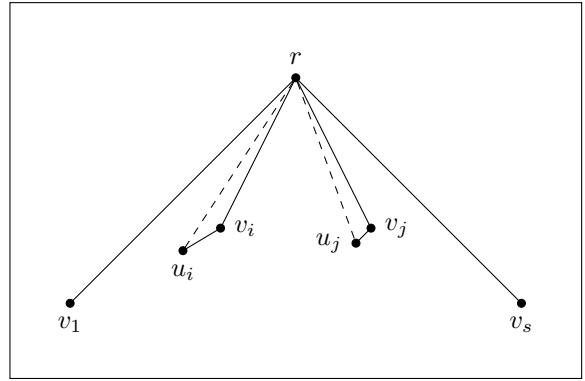
\begin{figure*}[htp]
\newcommand{\figymin}{-4}
\newcommand{\figymax}{1}
\newcommand{\figxmin}{-3.8}
\newcommand{\figxmax}{3.8}
\begin{center}
\subfigure[To have $u_j < v_i$ without edge crossings, $v_j u_j$ must meet $v_i v_s$ as shown.  But then, the three sides of the triangle $v_i v_j v_s$ cannot be blocked by the same convex obstacle $O_1$, which is a contradiction.]{
\begin{tikzpicture}[scale=1]
\path[draw = black] (\figxmin, \figymin) rectangle (\figxmax, \figymax);
\path (0,0) 			coordinate (r_coord) ;
\path (-3, -3) 		coordinate (v1_coord);
\path (3, -3) 		coordinate (vs_coord);
\path (-1, -2) 		coordinate (vi_coord);
\path (1, -2) 		coordinate (vj_coord);
\path (r_coord) 		node[black node] (r) 	[label=above:$r$] {};
\path (v1_coord) 	node[black node] (v1) [label=below:$v_1$] {};
\path (vi_coord) 		node[black node] (vi) 	[label=above left:$v_i$] {};
\path (vj_coord) 		node[black node] (vj) 	[label=right:$v_j$] {};
\path (vs_coord) 		node[black node] (vs) 	[label=below:$v_s$] {};
\draw (r_coord) -- (v1_coord);
\draw (r_coord) -- (vi_coord);
\draw (r_coord) -- (vj_coord);
\draw (r_coord) -- (vs_coord);
\draw[nonedge] (vi) -- (vj) -- (vs) -- (vi);
\path (-1.5, -2.4)		coordinate	(uj_coord);
\path (uj_coord) 		node[black node] (uj) 	[label=below:$u_j$] {};
\draw (vj_coord) -- (uj_coord);
\end{tikzpicture}
\label{fig:vi_lt_ui}
} 
\quad
\subfigure[Since $u_i < v_i < u_j < v_j$, no convex obstacle can meet both $ru_i$ and $ru_j$ without crossing an edge, so we have a contradiction to the assumption that \emph{three} non-edges of the form $ru_i$ can be blocked by the same obstacle $O'$.]{
\begin{tikzpicture}[scale=1]
\path[draw = black] (\figxmin, \figymin) rectangle (\figxmax, \figymax);
\path (0,0) 			coordinate (r_coord) ;
\path (-3, -3) 		coordinate (v1_coord);
\path (3, -3) 		coordinate (vs_coord);
\path (-1, -2) 		coordinate (vi_coord);
\path (1, -2) 		coordinate (vj_coord);
\path (r_coord) 		node[black node] (r) 	[label=above:$r$] {};
\path (v1_coord) 	node[black node] (v1) [label=below:$v_1$] {};
\path (vi_coord) 		node[black node] (vi) 	[label=right:$v_i$] {};
\path (vj_coord) 		node[black node] (vj) 	[label=right:$v_j$] {};
\path (vs_coord) 		node[black node] (vs) 	[label=below:$v_s$] {};
\draw (r_coord) -- (v1_coord);
\draw (r_coord) -- (vi_coord);
\draw (r_coord) -- (vj_coord);
\draw (r_coord) -- (vs_coord);
\path (-1.5, -2.3)		coordinate	(ui_coord);
\path (ui_coord) 		node[black node] (ui) 	[label=below:$u_i$] {};
\path (0.8, -2.2)		coordinate	(uj_coord);
\path (uj_coord) 		node[black node] (uj) 	[label=left:$u_j$] {};
\draw (vj_coord) -- (uj_coord);
\draw (vi_coord) -- (ui_coord);
\draw[nonedge] (r) -- (ui);
\draw[nonedge] (r) -- (uj);
\end{tikzpicture}
\label{fig:nothree}
} 
\end{center}
\caption{For the proof of Lemma \ref{lem:planeTreeUnbounded}.  Since all non-edges among $v_1, v_2, \ldots, v_s$ are blocked by a single convex obstacle $O_1$,  these vertices are in convex position and below $r$ in the manner shown in both subfigures.
} 
\label{fig:trappedPath}
\end{figure*}

\begin{lemma}
\label{lem:planeTreeUnbounded}
For every $m \in \ZZ^{+}$, there is a value of $k$ such that $T_{k, 2}$ has no $m$-convex obstacle representation without edge crossings.
\end{lemma}
\begin{proof}
Denote by $V_1$ the set of vertices at depth $1$, which is an independent set in $T_{k,2}$ of size $k$.
For any given $s$, we can find a subset $V' \subseteq V_1$ of size $s$ (provided large enough $k=k(s)$) such that
every non-edge with both endpoints in $V'$ is blocked by a common obstacle $O_1$.
This is because we can assign every non-edge among $V_1$ to a single obstacle that blocks it to obtain an $m$-edge-coloring of a $K_k$ induced on $V_1$,
which by Ramsey's Theorem has a monochromatic clique of size $s$ for large enough $k$.
The set $V'$ lies in some half-plane having $r$ on its boundary, without loss of generality, below a horizontal line;
otherwise, $r$ would be inside a triangle with vertices in $V'$, yet no single convex obstacle could block all three sides of it without meeting an edge of $T_{k,2}$.
Let us write $u < v$ whenever the triple $ruv$ is counterclockwise.
Let $v_1 < v_2 < \ldots < v_s$ denote the vertices in $V'$.
For each $i: 1 < i < s$, let $u_i$ denote a certain child of $v_i$.
We claim that at least $(s-2)/2$ (not necessarily disjoint) convex obstacles are required
to block the non-edges $ru_2, ru_3, \ldots, ru_{s-1}$.

To prove the claim, assume for contradiction that some obstacle $O'$ blocks three non-edges of the form $ru_i$.
Then without loss of generality, for some pair $i<j$ such that $ru_{i}$ and $ru_{j}$ are blocked by $O'$,
both $u_{i} < v_{i}$ and $u_{j} < v_{j}$ hold.
It must be that $v_{i} < u_{j}$;
otherwise, $v_j u_j$ would cross an edge 
or meet $O_1$ which blocks both $v_{i} v_{j}$ and $v_{j} v_{j+1}$.
See Figure \ref{fig:vi_lt_ui}.
Choose two points $p_i \in ru_i \cap O'$ and $p_j \in ru_j \cap O'$.
Then the segment $p_i p_j$ must intersect the union of the edges $r v_i$ and $v_i u_i$.
See Figure \ref{fig:nothree}.
By the convexity of $O'$, we have a contradiction.

Thus, for $s \geq 2m+4$, at least $m+1$ convex obstacles are required if no edges cross.
\end{proof}

Assume for contradiction that we have a representation of $T_{k,3}$ with \emph{three} pairwise disjoint convex obstacles $O_1$, $O_2$ and $O_3$.

If the endpoints of a crossing induced only the two edges (that is, an ``X" type crossing), at least four convex obstacles would be needed to block the non-edges, since any convex set that intersects two non-edges must meet an edge.
However, no more than three edges can be induced by four vertices without forcing a cycle.
Therefore, the four endpoints of every crossing induce a path with three edges.

By Lemma \ref{lem:planeTreeUnbounded},
we know that for large enough $k$, there are crossings within each subtree of $T_{k,3}$ isomorphic to $T_{k,2}$.
Pick three crossings $c_1, c_2,$ and $c_3$ in $T_{k,3}$, each in a subtree rooted at a different neighbor of the root vertex.
For each $i \in \{1,2,3\}$, denote by $u_{i} v_{i}$ and $w_{i} z_{i}$ the edges of the crossing $c_i$,
with the corresponding induced path on four vertices $P_i = u_{i} v_{i} w_{i} z_{i}$.

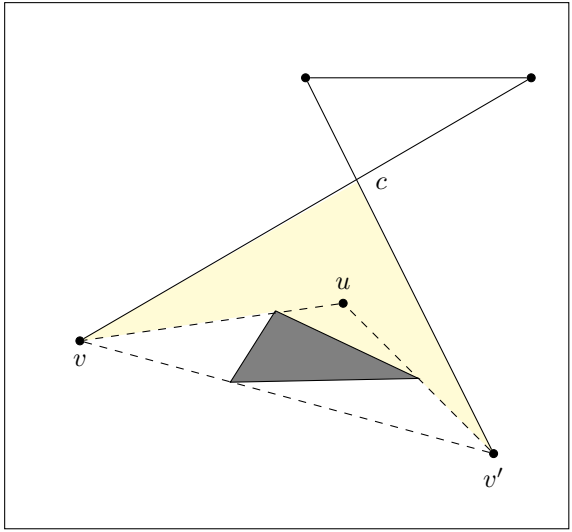
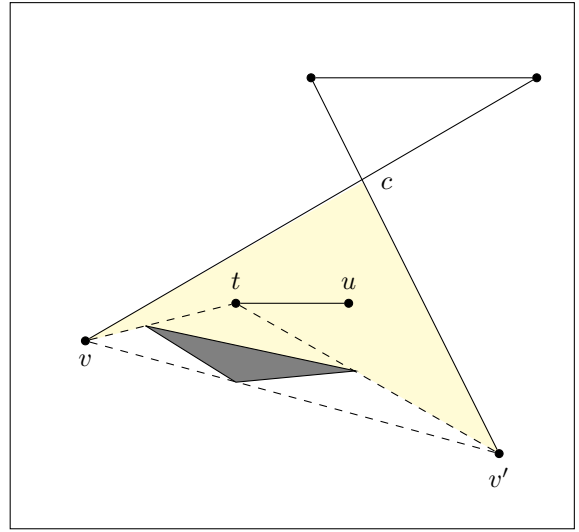
\begin{figure*}[htp]
\newcommand{\figymin}{-1}
\newcommand{\figymax}{6}
\newcommand{\figxmin}{0}
\newcommand{\figxmax}{7.5}
\begin{center}
\subfigure[Since the obstacle blocking $vv'$ must also be responsible for blocking $uv$ and $uv'$,
every neighbor of $u$ must be inside the lightly shaded region inside triangle $vcv'$.]{
\begin{tikzpicture}[scale=1]
\path[draw = black] (\figxmin, \figymin) rectangle (\figxmax, \figymax);

\path (4.68, 3.6) 	coordinate (c_coord) ;
\path (7, 5) 		coordinate (dummy1_coord);
\path (4,5) 			coordinate (dummy2_coord);
\path (1, 1.5) 		coordinate (v_coord);
\path (6.5, 0) 		coordinate (vprime_coord);
\path (4.5, 2) 		coordinate (u_coord);

\path (3.6, 1.9)			coordinate (ob1nodeuv_coord);
\path (5.5, 1)			coordinate (ob1nodeuvprime_coord);
\path (3, 0.95)			coordinate (ob1nodevvprime_coord);

\draw[carve] (vprime_coord) -- (c_coord)  -- (v_coord)  -- (ob1nodeuv_coord) -- (ob1nodeuvprime_coord);
\draw[carveob] (ob1nodeuv_coord) -- (ob1nodeuvprime_coord) -- (ob1nodevvprime_coord) -- (ob1nodeuv_coord);

\path (c_coord) 			node[white] (c) 		[label = right:$c$] {};
\path (dummy1_coord) 	node[black node] (dummy1) {};
\path (dummy2_coord) 	node[black node] (dummy2) {};
\path (v_coord) 			node[black node] (v) 		[label=below:$v$]{};
\path (vprime_coord) 		node[black node] (vprime) 	[label=below:$v'$]{};
\path (u_coord) 			node[black node] (u) 		[label=above:$u$] {};

\draw[edge] (dummy1) -- (v);
\draw[edge] (dummy2) -- (vprime);
\draw[nonedge] (v) -- (vprime);
\draw[edge] (dummy1) -- (dummy2);
\draw[nonedge] (vprime) -- (u) -- (v);
\end{tikzpicture}
\label{fig:firstVertex}
} 
\qquad
\subfigure[But every neighbor $t$ of $u$ is subject to the same conditions as $u$!  Hence, the neighbors of $t$ must be inside the lightly shaded region inside triangle $vcv'$\ldots]{
\begin{tikzpicture}[scale=1]
\path (4.68, 3.6) 	coordinate (c_coord) ;
\path (7, 5) 		coordinate (dummy1_coord);
\path (4,5) 			coordinate (dummy2_coord);
\path (1, 1.5) 		coordinate (v_coord);
\path (6.5, 0) 		coordinate (vprime_coord);
\path (4.5, 2) 		coordinate (u_coord);

\path (1.8, 1.7)			coordinate (ob1nodetv_coord);
\path (4.59, 1.1)			coordinate (ob1nodetvprime_coord);
\path (3, 0.95)			coordinate (ob1nodevvprime_coord);

\path (c_coord) 			node[white] (c) 		[label = right:$c$] {};

\draw[carve] (vprime_coord) -- (c_coord)  -- (v_coord)  -- (ob1nodetv_coord) -- (ob1nodetvprime_coord);

\path (dummy1_coord) 	node[black node] (dummy1) {};
\path (dummy2_coord) 	node[black node] (dummy2) {};
\path (v_coord) 			node[black node] (v) 		[label=below:$v$]{};
\path (vprime_coord) 		node[black node] (vprime) 	[label=below:$v'$]{};
\path (u_coord) 			node[black node] (u) 		[label=above:$u$] {};
\path (3, 2) node[black node] (t) [label=above:$t$] {};

\draw[carveob] (ob1nodetv_coord) -- (ob1nodetvprime_coord) -- (ob1nodevvprime_coord) -- (ob1nodetv_coord);

\draw[edge] (dummy1) -- (v);
\draw[edge] (dummy2) -- (vprime);
\draw[nonedge] (v) -- (vprime);
\draw[edge] (dummy1) -- (dummy2);
\draw[nonedge] (v) -- (t) -- (vprime) ;
\draw[edge] (t) -- (u);

\path[draw = black] (\figxmin, \figymin) rectangle (\figxmax, \figymax);
\end{tikzpicture}
\label{fig:anotherVertex}
} 

\end{center}
\caption{Non-edges shown in each subfigure imply a respective minimal portion (dark gray) of an obstacle.
The third edge of the path could have been incident on $v$ or $v'$ but this makes no difference.  Only the obstacle that blocks $vv'$ can be inside the convex angle $v'cv$.} 
\label{fig:trappedPath}
\end{figure*}

Let us first consider the case where the convex hulls of two of these paths, say $P_1$ and $P_2$, meet.
If this is the case, with no vertex of $P_1$ being inside the convex hull of $P_2$ or vice versa, 
then some edge of $P_1$ must intersect some edge of $P_2$,
inducing an ``X'' type crossing which requires four obstacles.
Hence, without loss of generality, some vertex $u$ of $P_1$ is in the convex hull of $P_2$.
See Figure \ref{fig:trappedPath}.
Let $c$ be the point of intersection of the two edges of $P_2$.
Then, $u$ is inside some triangle $v c v'$ where $v, v' \in P_2$.
If $vv'$ is an edge, then $vcv'$ induces a bounded face, so $uv$ would require an obstacle in addition to the three required by $P_2$.
Now, since $u$ is inside a triangle $vcv'$, the obstacle blocking $vv'$ must also block $uv$ and $uv'$, but this forces all neighbors of $u$ to be inside $vcv'$.
Applying this argument to the neighbors of $u$ in $P_1$ (recursively if needed), which satisfy the same conditions as $u$, we see that $P_1$ must be completely inside $vcv'$.
But every non-edge of $P_1$ requires a distinct obstacle, at most one of which may coincide with one blocking $vv'$ while none among them may coincide with any other obstacle, so five obstacles are required, a contradiction.

\begin{figure}[htp]
\centering
\includegraphics[scale=0.8]{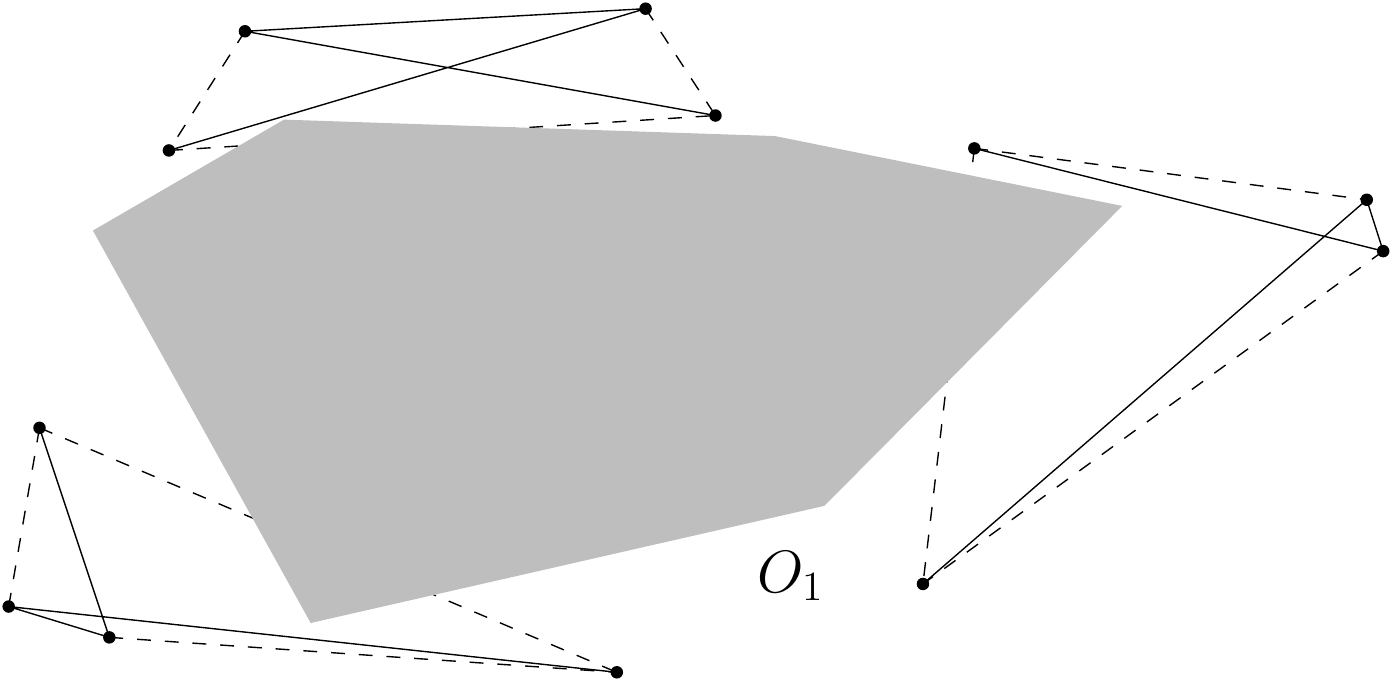}
\caption{convex hulls of $P_1$, $P_2$ and $P_3$ are pairwise disjoint}
\label{fig:3path2}
\end{figure}

This means that the convex hulls of $P_1$, $P_2$ and $P_3$ are pairwise disjoint.
Recall that for each of these paths $P_i$, each of the three non-edges of $P_i$ must be blocked by a unique obstacle among three pairwise disjoint obstacles.
Hence by the Jordan Curve Theorem we get a contradiction (see Figure \ref{fig:3path2}).
\end{proof}

\section{Convex obstacle number of bipartite permutation graphs}
\label{sec:bipartite_permutation_upper_bound_4}
A \emph{permutation graph} is a graph on $[n]$ according to a permutation $(\sigma_1, \sigma_2, \ldots, \sigma_n)$ of $[n]$ such that there is an edge between two elements $\sigma_i > \sigma_j$ 
whenever
$i < j$.
We show that the idea of having a small aperture between two classes of vertices, which are placed close to two convex obstacles, is readily extended to the class of bipartite permutation graphs.

\begin{proof}[Proof of Theorem \ref{thm:bipartite_permutation_upper_bound_4}]
By a result from \cite{permutationGraph}, a bipartite graph $G(V,E)$ is a permutation graph if and only if 
its two independent vertex classes $V_1$ and $V_2$ can be ordered such that the neighborhood of every vertex $u_i \in V_1$ forms an interval $[a_i, b_i]$ in $V_2$, and if $u_i < u_j$ for two vertices in $V$ then
$a_i \leq a_j$ and $b_i \leq b_j$.

We illustrate in Figure \ref{fig:permutation} a set $\mathcal{C}$ of \emph{four} disjoint convex obstacles allowing an obstacle representation of $G$.  $\mathcal{C}$ consists of two convex arcs $C_1$ and $C_2$,
and two vertical line segments (labeled $\mathcal{A}$) which form an aperture between $C_1$ and $C_2$.

\begin{figure}[htp]
\centering
\begin{tikzpicture}[scale=1.2]
\newcommand{\aby}{0.5}
\newcommand{\dx}{2};
\newcommand{\dyfora}{1.9};
\newcommand{\dyforb}{2.05};
\coordinate (a) at (0,0);
\coordinate (b) at ($(a) + (0,\aby)$);
\coordinate (u_{i+1}) at ($(a) + (-\dx,\dyfora)$);
\coordinate (u_i) at ($(b) + (-\dx,\dyforb)$);
\coordinate (a_{i+1}) at ($(a) + (\dx,-\dyfora)$);
\coordinate (b_i) at ($(b) + (\dx,-\dyforb)$);
\coordinate (b_{i+1}) at ($(a_{i+1}) + (0,{2*\aby})$);
\coordinate (a_i) at ($(b_i) - (0,2*\aby)$);

\node[black node] (a) at (a) [label=below left:$a$] {};
\node[black node] (b) at (b) [label=above right:$b$] {};
\node[black node] (u_{i+1}) at (u_{i+1}) [label=left:$u_{i+1}$] {};
\node[black node] (u_i) at (u_i) [label=left:$u_i$] {};
\node[black node] (a_{i+1}) at (a_{i+1}) [label=right:$a_{i+1}$] {};
\node[black node] (b_i) at (b_i) [label=right:$b_i$] {};
\node[black node] (a_i) at (a_i) [label=right:$a_i$] {};
\node[black node] (b_{i+1}) at (b_{i+1}) [label=right:$b_{i+1}$] {};

\draw (u_{i+1}) -- (a_{i+1});
\draw (u_i) -- (b_i);
\draw (u_{i+1}) -- (b_{i+1});
\draw (u_i) -- (a_i);

\newcommand{\topcoord}{3}
\newcommand{\bottomcoord}{-3}
\newcommand{\rightcoord}{\dx}
\draw[very thick] (a) -- (0, \bottomcoord);
\draw[very thick] (b) -- (0, \topcoord);
\draw[very thick] (\rightcoord, \topcoord) -- (\rightcoord, \bottomcoord);
\draw[very thick] (-\rightcoord, \topcoord) -- (-\rightcoord, \bottomcoord);
\node (C_1) at (-\dx,0) [label=left: \large{$C_1$}]{};
\node (C_2) at (\dx,0) [label=right: \large{$C_2$}]{};
\end{tikzpicture}
\caption{The obstacles allowing a obstacle representation of a bipartite permutation graph.}
\label{fig:permutation}
\end{figure}
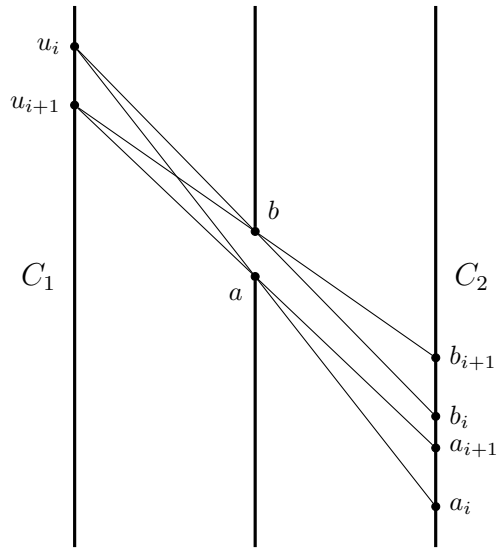

Similar to the treatment of the arcs in Subsection \ref{sec:up:5-convex}, we regard $C_1$ and $C_2$ as line segments, except that they block visibilities among graph vertices placed near them. For convenience, we shall speak of placing vertices of $G$ \emph{on} these arcs.

We put vertices of $V_1$ and $V_2$ on $C_1$ and $C_2$ respectively. Let $u_1,u_2, \ldots, u_n$ and $v_1,v_2, \ldots, v_n$ be the ordering of the vertices in $V_1$ and $V_2$ guaranteed by the aforementioned result in \cite{permutationGraph}. We place the vertices, in order, inductively. In the basis step, we place $u_1$ arbitrarily on the relative interior of $C_1$. Let $a_i$ and $b_i$ denote the endpoints of the segment of $C_2$ that $u_i$ can see through the aperture (see Figure \ref{fig:permutation}). We place neighbors of $u_1$ in the relative interior of segment $a_{1}b_{1}$ on $C_2$ so that the order of their $y$-coordinates corresponds to their order in $V_2$.

At an inductive step $i+1$, where $i\geq 1$, we place the $(i+1)$-th vertex of $V_1$ together with its children, on the corresponding arcs as follows. We first find a consistent place for $a_{i+1}$. If the first neighbor $w$ of the $(i+1)$-th vertex (with regards to the order in $V_2$) is already placed on $C_2$, we pick $a_{i+1}$ so that it precedes $w$ (with regards to $y$-coordinate) and succeeds $a_i$ and any other point already placed on $C_2$. Otherwise, we pick $a_{i+1}$ so that it succeeds $b_i$. We place $u_{i+1}$ at the intersection of $C_1$ and the line through $a_{i+1}$ and $a$ (see Figure~\ref{fig:permutation}). The line through $u_{i+1}$ and $b$ intersects $C_2$ at point $b_{i+1}$, which has a higher $y$-coordinate than $b_{i}$. Therefore we can place neighbors of $u_{i+1}$ that are not neighbors of $u_i$ on the non-empty line segment $b_{i}b_{i+1}$.\
This concludes the proof of Theorem \ref{thm:bipartite_permutation_upper_bound_4}.
\end{proof}

\subsection*{Acknowledgments}
The authors are indebted to Filip Mori{\'{c}} for fruitful discussions
and an idea that led to the proof of our upper bound construction for outerplanar graphs.
The authors would also like to thank the anonymous referee and Boris Aronov for providing useful feedback.

\end{document}